\documentclass[preprint,12pt]{elsarticle}

\usepackage{amssymb}
\usepackage{amsmath}

\journal{Nucl. Instr. and Meth. A}

\begin{document}

\begin{frontmatter}



\title{Improvement of the energy resolution of CdTe detectors by pulse height correction from waveform}


\author{T.Kikawa}
\author{A.K.Ichikawa}
\author{T.Hiraki}
\author{T.Nakaya}
\address{Department of Physics, Kyoto University, Kyoto 606-8502, Japan}
\begin{abstract}
Semiconductor detectors made of CdTe crystal have high $\gamma$-ray detection efficiency and are usable at room temperature.
However, the energy resolution of CdTe detectors for MeV $\gamma$-rays is rather poor because of the significant hole trapping effect.
We have developed a method to improve the energy resolution by correcting the pulse height using the waveform of the signal.
The energy resolution of 2.0\% (FWHM) for 662keV $\gamma$-rays
was achieved with an ohmic-contact 5.0mm $\times$ 5.0mm $\times$ 5.0mm crystal.
Best energy resolution was achieved at temperatures between ${-10}^\circ\mathrm{C}$ and ${0}^\circ\mathrm{C}$.

\end{abstract}


\end{frontmatter}


\section{Introduction}
Semiconductor detectors made of cadmium telluride (CdTe) have been developed and used as radiation detectors having high $\gamma$-ray
detection efficiency.
This high efficiency is attributed to its high density (5.85g/cm$^3$) and large atomic number ($Z_{Cd}$=48, $Z_{Te}$=52).
The relatively high band gap energy (1.53eV) makes the detector usable at room temperature.
However, the energy resolution of CdTe detectors for MeV $\gamma$-rays is rather poor ($\sim$10\% FWHM).
This is mainly because generated holes are trapped while they are drifting.
When there is significant hole trapping, the collected charge in a semiconductor detector having homogeneous electric field is expressed as follows:
\begin{equation}
Q=\frac{Ne}{d}\left(x_0+\frac{V\mu_h\tau}{d}\left(1-\exp\frac{-d\left(d-x_0\right)}{V\mu_h\tau}\right)\right),\label{shiki}\\
\end{equation}
where $N$ is the number of generated carrier pairs, $\mu_h$ the mobility of holes,
$\tau$ life time of holes,
$V$ the applied voltage,
$d$ the distance between the electrodes,
and $x_0$ the position of carrier generation from the anode.
In comparison, the collected charge in the Ge detectors depends only on the number of generated carrier pairs
because $\mu_h\tau$ is large and the effect of hole trapping is negligibly small.
In case of CdTe detectors, the collected charge depends on the position of carrier generation
because $\mu_h\tau$ is small (7$\times 10^5$cm$^2$/V at 300K\cite{cdte_mobility}) and the effect
of hole trapping is noticeable. It degrades the energy resolution.
Devices thinner than 2mm are practically used as X-ray or low-energy $\gamma$-ray detectors
because the effect of hole trapping is small with such thin devices.
For MeV $\gamma$-rays, such thin devices not only have low detection efficiency
but also can not measure the energy precisely
because, in many cases, an electron generated with the photoelectric absorption escapes from the device.
To achieve high energy resolution for MeV $\gamma$-rays with thick crystals,
methods of cancelling the signal induced by the hole drift using specially structured electrodes\cite{cpg}\cite{pixel}\cite{capture}
have been developed. They are used in experiments using CdTe detectors or CdZnTe
detectors\footnote{Characteristics of CdZnTe detectors are similar to CdTe detectors.}
and demonstrated high energy resolution with thick crystals ($\gtrsim$1\% FWHM for 662keV $\gamma$-rays).
On the other hand, methods of correcting the pulse signal from simple planar electrode devices have also been
developed to achieve high energy resolution.
However, high energy resolution with these methods has been demonstrated so far only for thin crystals
up to 2mm in thickness\cite{system}\cite{bargholtz}\cite{takahashi}\cite{abbene}, or for low energy $\gamma$-rays
for which only shallow region of the crystal was used\cite{montemont}.
In this paper, we demonstrate high energy resolution with rather thick (5mm) CdTe crystals by correcting the pulse signal.
Our method used the measured relation between the pulse height and the drift time.
Therefore, it does not rely on the specific models of charge transportation.

\section{Method of pulse height correction}
\begin{figure}[h]
\begin{minipage}[t]{15.2pc}
\includegraphics[width=15.2pc]{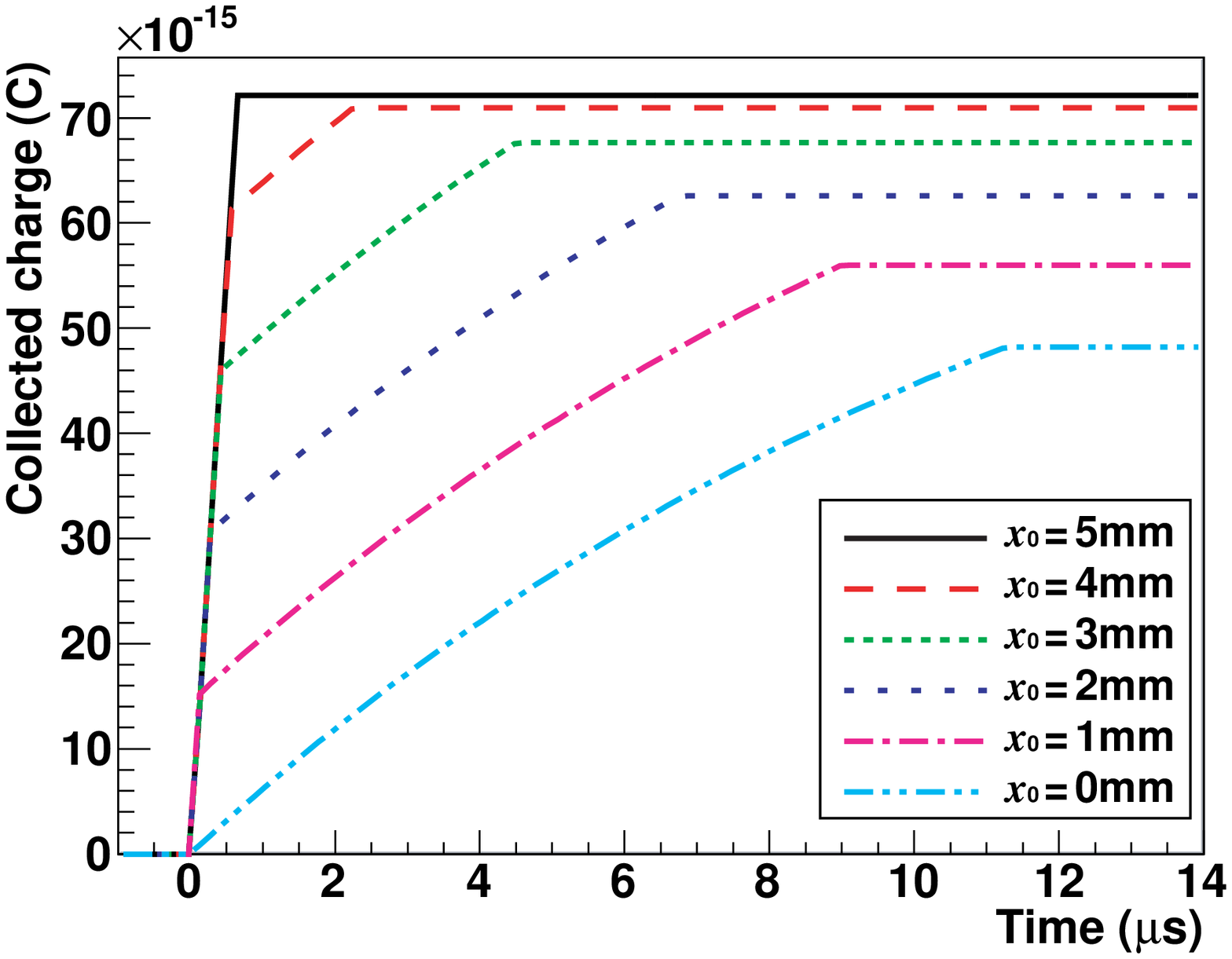}
\caption{\label{wave_theory}Expected waveform when $d$=5mm, $V$=400V, $E_{\gamma}$=662keV, $\mu_e$=880cm$^2$/(V$\cdot$s), $\mu_h$=90cm$^2$/(V$\cdot$s), $\tau$=7.8$\times$10$^{-7}$s\cite{cdte_mobility}}
\end{minipage}\hspace{1.9pc}%
\begin{minipage}[t]{15.2pc}
\includegraphics[width=15.2pc]{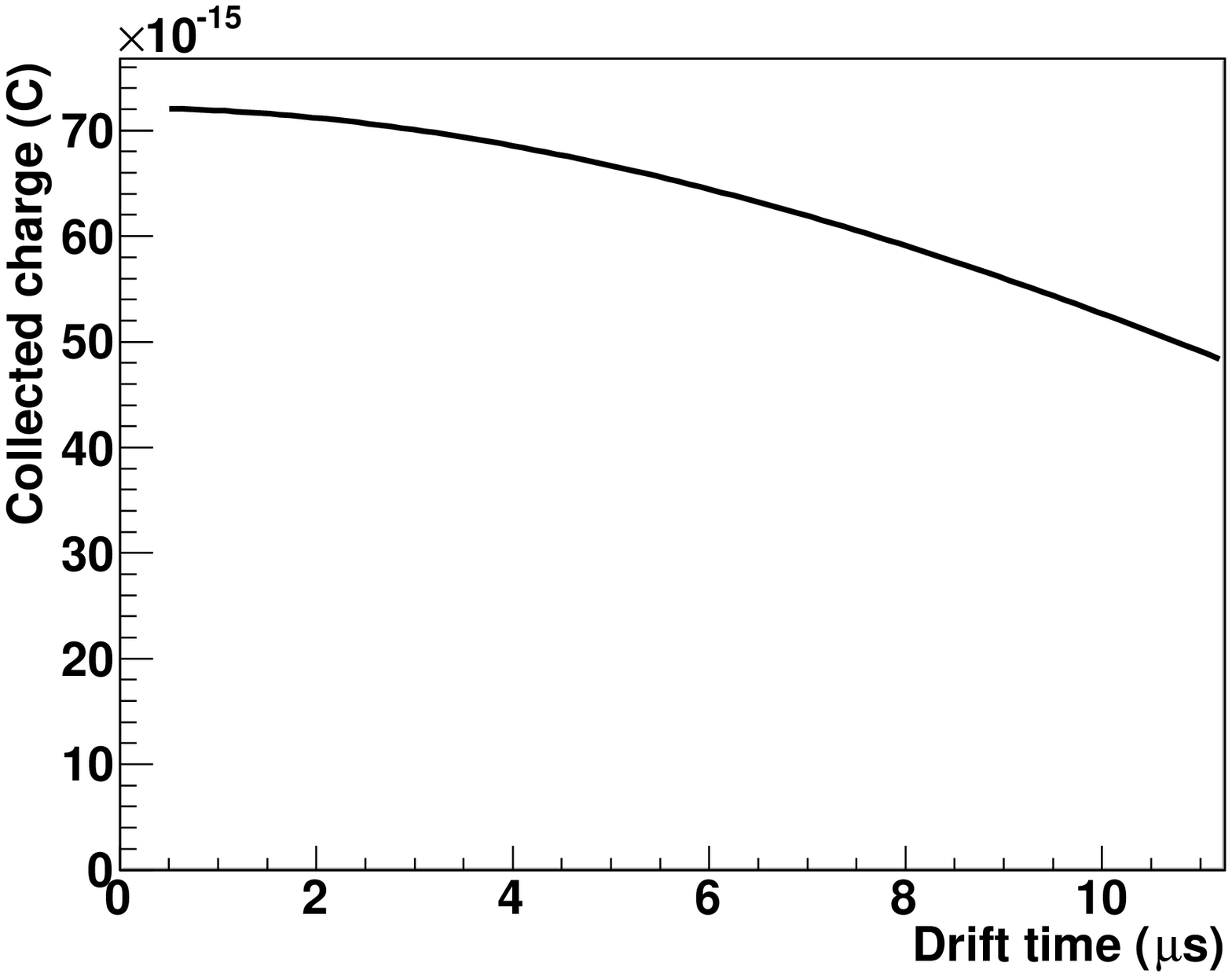}
\caption{\label{height_time_theory}Expected relation between the pulse height and the drift time when $d$=5mm, $V$=400V, $E_{\gamma}$=662keV, $\mu_e$=880cm$^2$/(V$\cdot$s), $\mu_h$=90cm$^2$/(V$\cdot$s), $\tau$=7.8$\times$10$^{-7}$s\cite{cdte_mobility}}
\end{minipage} 
\end{figure}

We have developed a method of correcting pulse signal to achieve high energy resolution with thick CdTe crystals.
In this method, the pulse height of the signal from the ohmic-contact CdTe detectors is corrected using the waveform information.
The electric field in the planar ohmic-contact semiconductor is nearly homogeneous.
Then, the waveform of the signal is expected to be as follows and Fig.\ref{wave_theory}:

\begin{align}
&Q(t)=\begin{cases}
\frac{NeV}{d^2}\left(\mu_et+\mu_h\tau\left(1-\exp(\frac{-t}{\tau})\right)\right) & \text{$\left(t<\frac{x_0d}{V\mu_e}\right)$} \\
\frac{Ne}{d}\left(x_0+\frac{V\mu_h\tau}{d}\left(1-\exp(\frac{-t}{\tau})\right)\right) & \text{$\left(\frac{x_0d}{V\mu_e}\leq t<\frac{(d-x_0)d}{V\mu_h}\right)$,} \\
\label{eq_pulse}
\frac{Ne}{d}\left(x_0+\frac{V\mu_h\tau}{d}\left(1-\exp(\frac{-d(d-x_0)}{V\mu_h\tau})\right)\right) & \text{$\left(t\geq \frac{(d-x_0)d}{V\mu_h}\right)$}
\end{cases}
\end{align}
where $t$ is the time from the carrier generation and $\mu_e$ the mobility of electrons.
A prompt rise followed by a slow rise can be seen in Fig.\ref{wave_theory}.
The former is mainly caused by the electron drift and the latter by the hole drift.
The effect of electron (hole) drift is large when carriers are generated near cathode (anode).
From the waveform, the drift time of the carriers, which is defined as a time from the leading to rear edges of
the rising signal pulse, can be estimated.
When the drift time is long, the pulse height is expected to be lower
because the effect of the hole trapping become more noticeable (Fig.\ref{height_time_theory}).
If signals from the CdTe detectors are read out by a flash ADC and the drift time is estimated from the obtained waveform,
the decline in the pulse height by the hole trapping can be estimated and corrected.
By this correction, it is expected that better energy resolution can be achieved.

\section{Measurement}
\begin{figure}[h]
\begin{center}
\includegraphics[width=26.0pc]{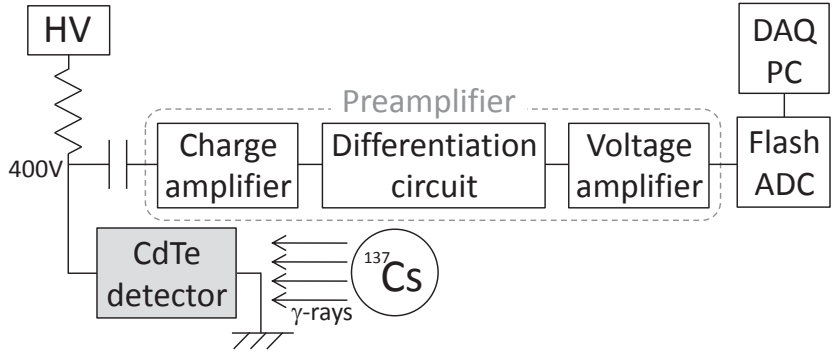}
\caption{\label{setup}The schematic diagram of the setup for the measurement}
\end{center}
\end{figure}

To demonstrate this method for MeV $\gamma$-rays,
measurements were done with a following setup.
An ohmic-contact CdTe detector with a 5.0mm $\times$ 5.0mm $\times$ 5.0mm Cl-doped monocrystal device
(CLEAR-PULSE$^{\textregistered}$ CdTe505050) was used.
A bias voltage of 400V was applied. The signal is processed by a charge-sensitive preamplifier (CLEAR-PULSE$^{\textregistered}$ 580K).
The preamplifier consists of a charge amplifier, a differentiation circuit and a voltage amplifier.
In order to avoid the ballistic deficit and to maintain the raw pulse shape,
the time constant of the differentiation circuit was changed from 60$\mu$s to 600$\mu$s.
The signal from the preamplifier is read out by a 100MHz flash ADC (CAEN$^{\textregistered}$ V1724).
The 662keV $\gamma$-rays from $^{137}$Cs are irradiated to the CdTe detector
from the cathode side.
Because the attenuation length of 662 keV $\gamma$-rays in CdTe is 2.3cm,
the number of interactions in the proximity of anode is about 20\% less than that in the proximity of cathode.
The data are triggered by the self trigger of the flash ADC module.
The schematic diagram of the setup is shown in Fig.\ref{setup}.
The temperature was held at ${0}^\circ\mathrm{C}$ unless otherwise noted.

\section{Result}
\subsection{Waveform}

\begin{figure}[h]
\begin{center}
\includegraphics[width=15.2pc]{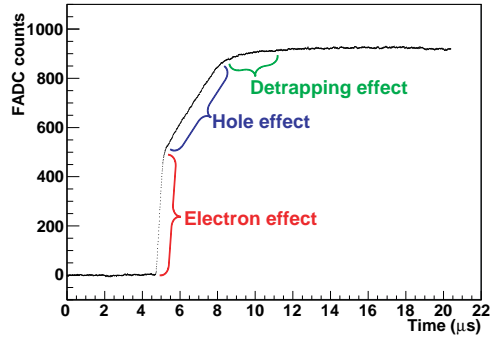}
\caption{\label{waveform}Typical waveform of the signal from the CdTe detector}
\end{center}
\end{figure}

\begin{figure}[h]
\begin{minipage}[t]{15.2pc}
\includegraphics[width=15.2pc]{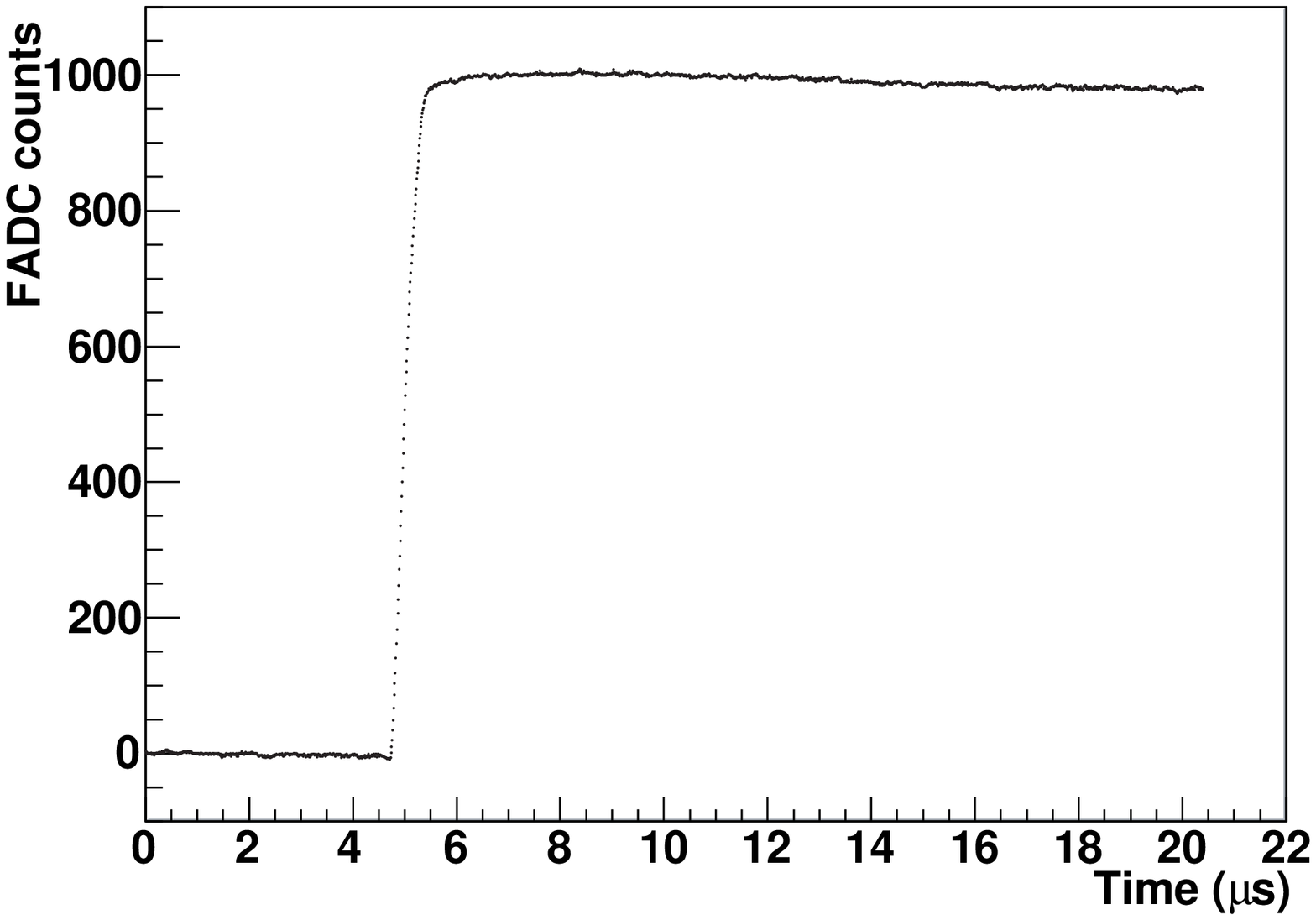}
\caption{\label{anode}Example waveform dominated by the prompt rise}
\end{minipage}\hspace{1.9pc}%
\begin{minipage}[t]{15.2pc}
\includegraphics[width=15.2pc]{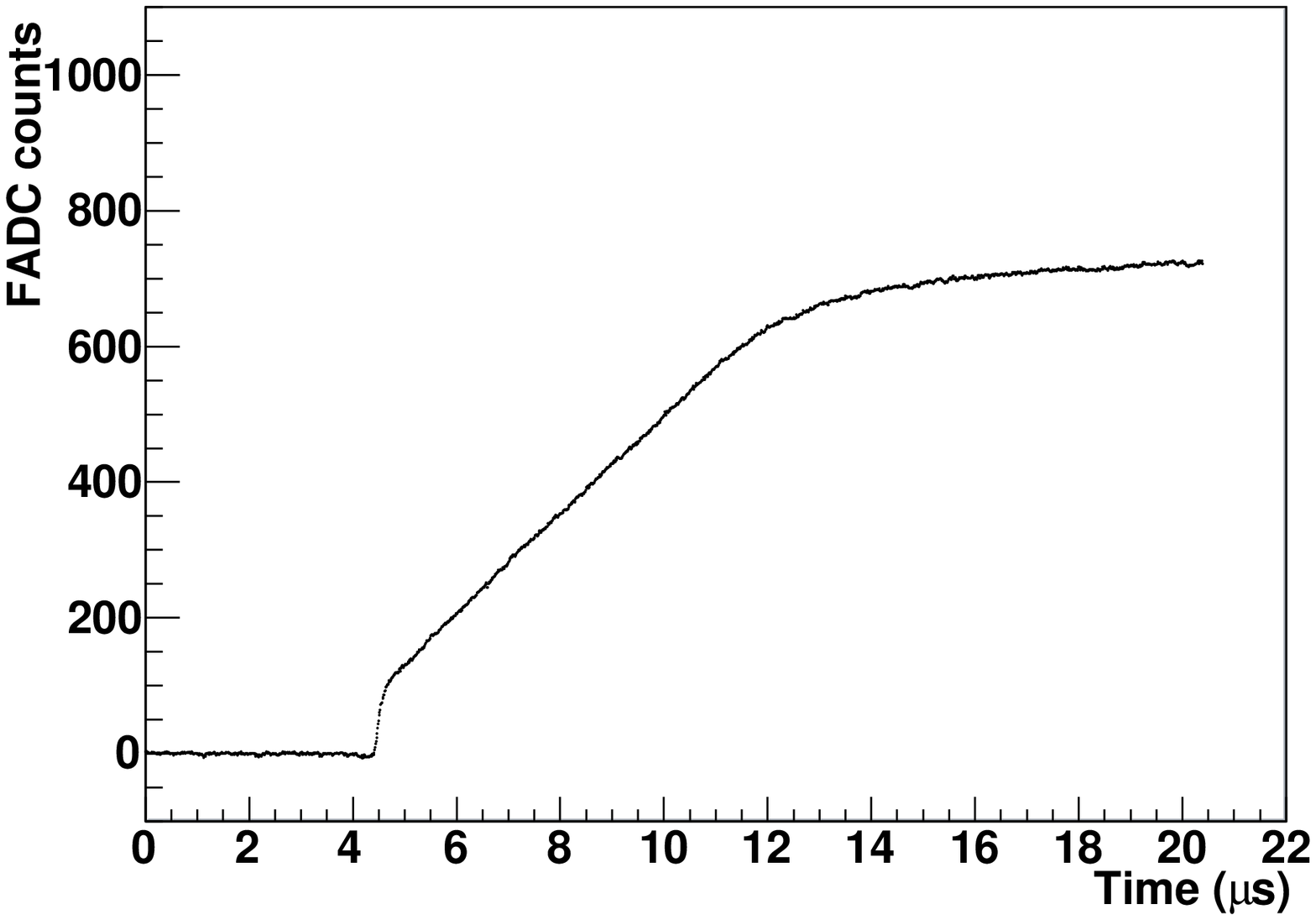}
\caption{\label{cathode}Example waveform dominated by the slow rise}
\end{minipage} 
\end{figure}

Figure \ref{waveform} is a typical observed waveform of the signal from the CdTe detector.
As expected, the prompt rise caused by the electron drift and the slow rise caused by the hole drift can be seen.
There are events in which the prompt rise is dominant as shown in Fig.\ref{anode}
and events where the slow rise is dominant as shown in Fig.\ref{cathode}.
The former are considered to be the events in which carriers were generated near the cathode and the latter
carriers were generated near the anode.
In Fig.\ref{waveform} and Fig.\ref{cathode}, the waveform keeps rising rather slowly even after the slow rise.
This may be caused by the release of shallowly-trapped holes (detrapping)\cite{detrapping}.

\subsection{Pulse height correction}

\begin{figure}[h]
\begin{center}
\includegraphics[width=18.3pc]{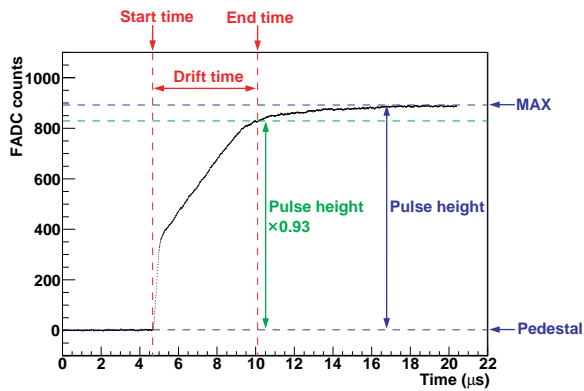}
\caption{\label{ana}Analysis method}
\end{center}
\end{figure}

The pulse height and the drift time are calculated from the recorded waveform.
The pulse height is defined as the difference between the highest flash ADC counts and the pedestal level.
The drift start time is taken as the time when the signal exceeds the pedestal level,
the drift end time as the time when the signal exceeds 93\% level of the pulse height,
and the drift time as difference between them (Fig.\ref{ana}).
This 93\% level is optimized to reduce the effect of the detrapping.

\begin{figure}[h]
\begin{minipage}[t]{15.2pc}
\includegraphics[width=15.2pc]{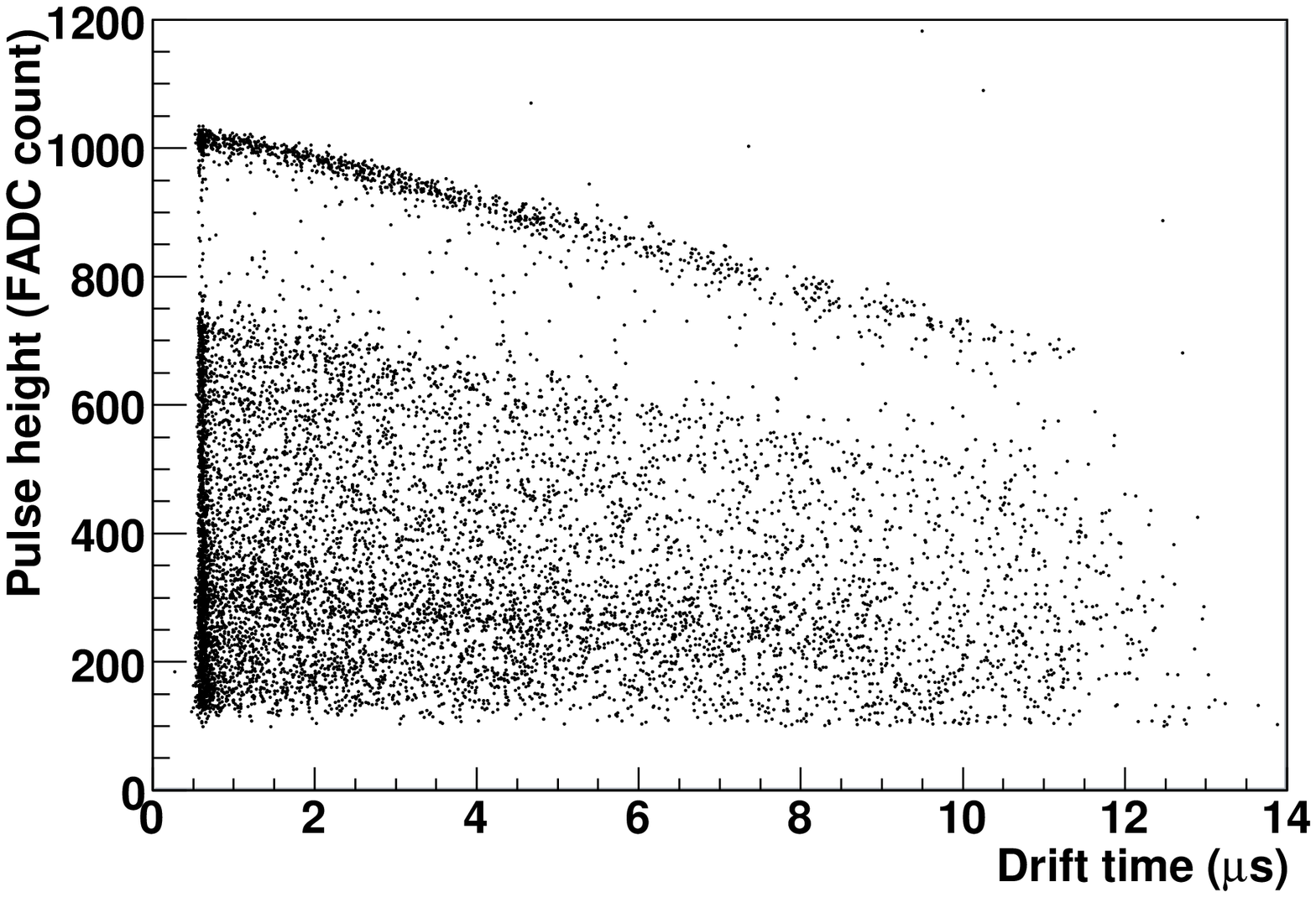}
\caption{\label{height_time}Distribution of the drift time and the pulse height from 662keV $\gamma$-rays}
\end{minipage}\hspace{1.9pc}%
\begin{minipage}[t]{15.2pc}
\includegraphics[width=15.2pc]{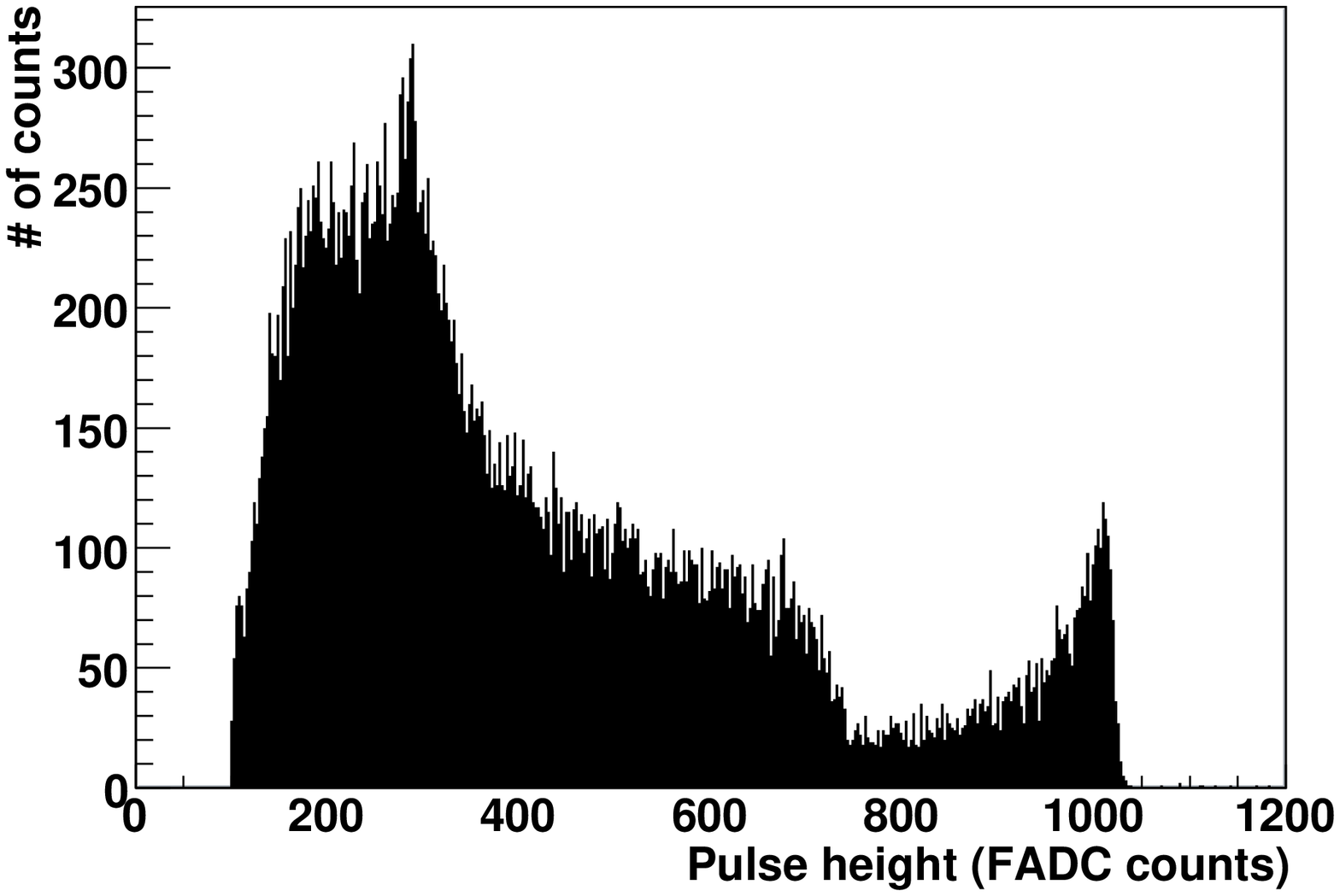}
\caption{\label{height}Distribution of the pulse height from 662keV $\gamma$-rays}
\end{minipage} 
\end{figure}

\begin{figure}[h]
\begin{minipage}[t]{15.2pc}
\includegraphics[width=15.2pc]{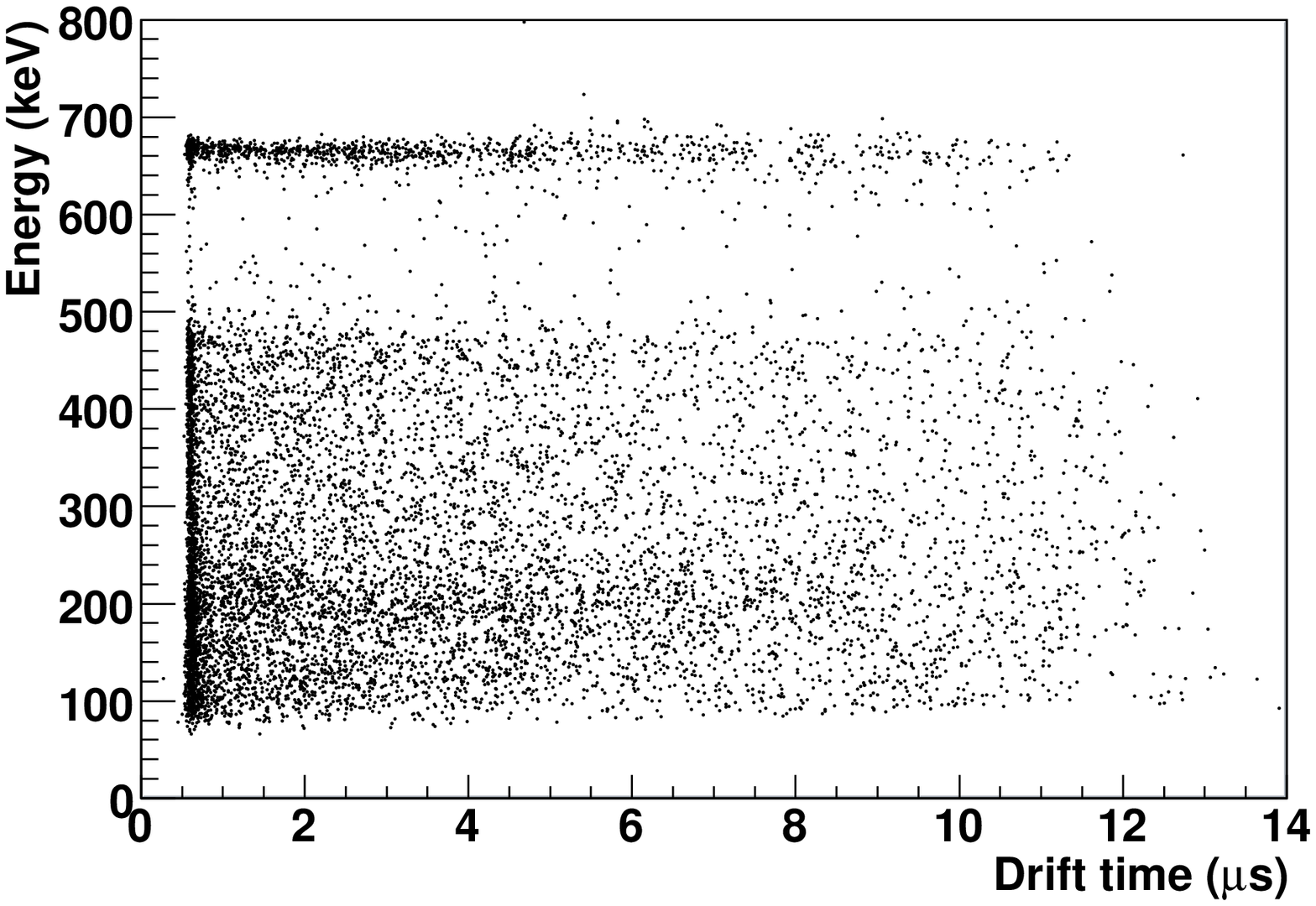}
\caption{\label{energy_time}Distribution of the drift time and the energy deposit by 662keV $\gamma$-rays estimated by the pulse height correction}
\end{minipage}\hspace{1.9pc}%
\begin{minipage}[t]{15.2pc}
\includegraphics[width=15.2pc]{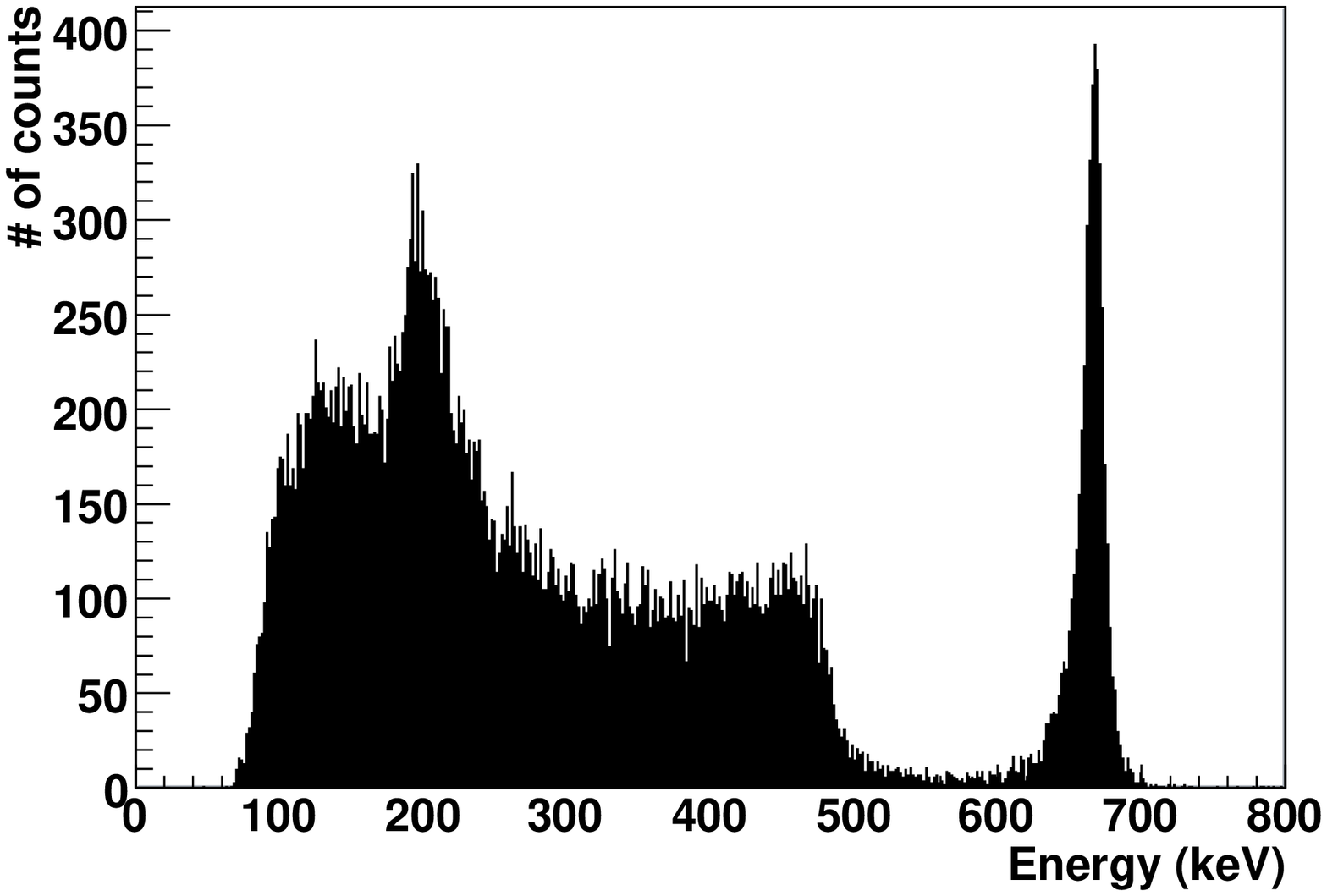}
\caption{\label{energy}Distribution of the energy deposit by 662keV $\gamma$-rays estimated by the pulse height correction}
\end{minipage} 
\end{figure}

The distribution of the pulse height and the drift time calculated in this way is shown in Fig.\ref{height_time}.
Figure \ref{height} shows the pulse height distribution.
Photoelectric absorption events and Compton scattering events can be identified in Fig.\ref{height_time}.
When the drift time is long, the pulse height becomes lower as expected.
This effect adds a large tail to the photoelectric absorption peak in the pulse height distribution (Fig.\ref{height}).
The relation between the pulse height and the drift time is determined
by fitting the photoelectric absorption part in Fig.\ref{height_time} with a cubic function.
The energy deposit in the CdTe device is estimated by correcting the pulse height from the drift time using this relation.
By this correction, the effect of the ballistic deficit (1.7\% when the drift time is 10$\mu$s) is also corrected.
The distribution of the estimated energy deposit and the drift time is shown in Fig.\ref{energy_time}
and the estimated energy deposit distribution in Fig.\ref{energy}.
The estimated energy deposit is no longer dependent on the drift time and
the photoelectric absorption peak in Fig.\ref{energy} is sharp.
The energy resolution is 2.0\% (FWHM).

\subsection{Consideration}
The energy resolution of semiconductor detectors $W_T$ consists of
the contribution from the statistical fluctuation of the number of generated carriers $W_D$,
incompleteness of carrier collection $W_X$ and
noise $W_E$:

\begin{equation}
{W_T}^2={W_D}^2+{W_X}^2+{W_E}^2.\label{bunkaino}
\end{equation}

$W_D$ of CdTe detectors for 662keV $\gamma$-rays is 0.24\%\footnote{$W_D$ can be calculated as follows\cite{wd}:
\begin{equation}
W_D=2.35\sqrt{\frac{F\varepsilon}{E_\gamma}},
\end{equation}
where $F$ is the Fano factor (assumed to be 0.15\cite{fano}) and $\varepsilon$ the energy to generate one electron-hole pair (4.43eV for CdTe).}.
$W_E$ was measured to be 0.79\% by
inputting signals generated by an accurate pulse generator
to the preamplifier with CdTe detector
and measuring the fluctuation of the signals from the preamplifier.
$W_X$ is expected to be 1.82\% from Eq.\ref{bunkaino}, and
it is still the main contribution even after the pulse height correction.
The source of $W_X$ might be fluctuation of basic properties such as electric field strength,
hole lifetime, detrapping time scale etc due to the non-uniformity of the crystal.

\section{Temperature dependence}
\begin{figure}[h]
\begin{center}
\includegraphics[width=15.2pc]{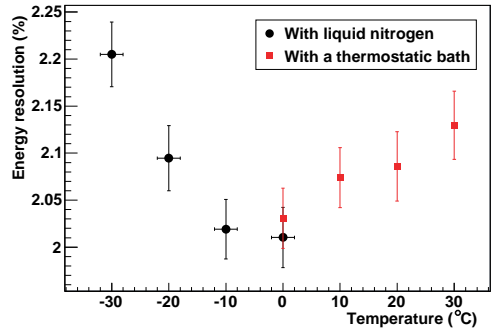}
\caption{\label{resolution_temperature}Obtained energy resolutions at various temperatures. The error bars are statistical.}
\end{center}
\end{figure}

The temperature dependence of the performance was measured.
Temperature of the CdTe device was controlled in the range from ${-80}^\circ\mathrm{C}$ to ${0}^\circ\mathrm{C}$ with liquid nitrogen cooling
and from ${0}^\circ\mathrm{C}$ to ${30}^\circ\mathrm{C}$ with a thermostatic bath.
The temperature was monitored by a platinum thermometer during the liquid nitrogen cooling.
Figure \ref{resolution_temperature} shows the obtained energy resolution at various temperatures.

\begin{figure}[h]
\begin{minipage}[t]{15.2pc}
\includegraphics[width=15.2pc]{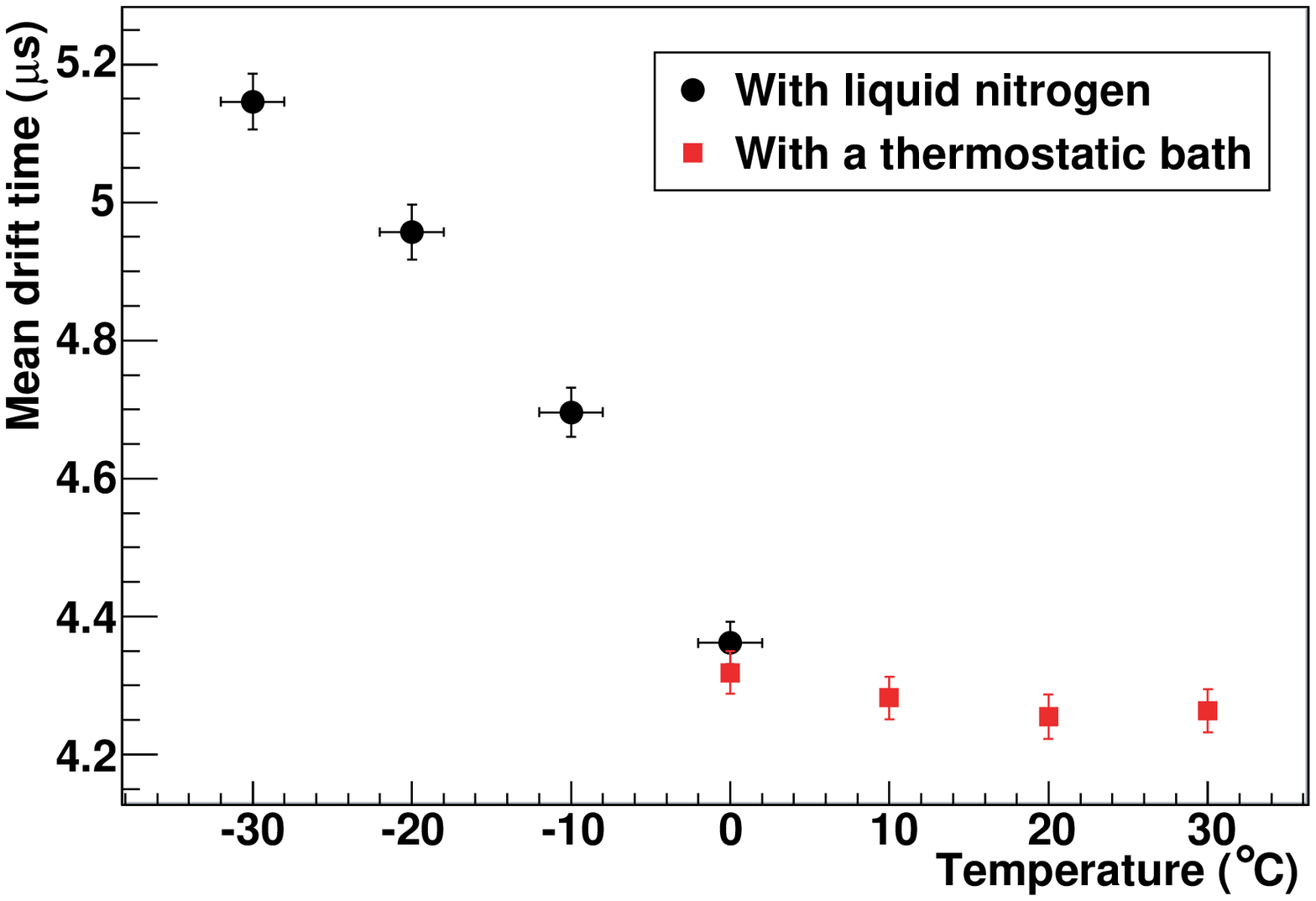}
\caption{\label{drift_time_temperature}Mean drift time at various temperatures. The error bars are statistical.}
\end{minipage}\hspace{1.9pc}%
\begin{minipage}[t]{15.2pc}
\includegraphics[width=15.2pc]{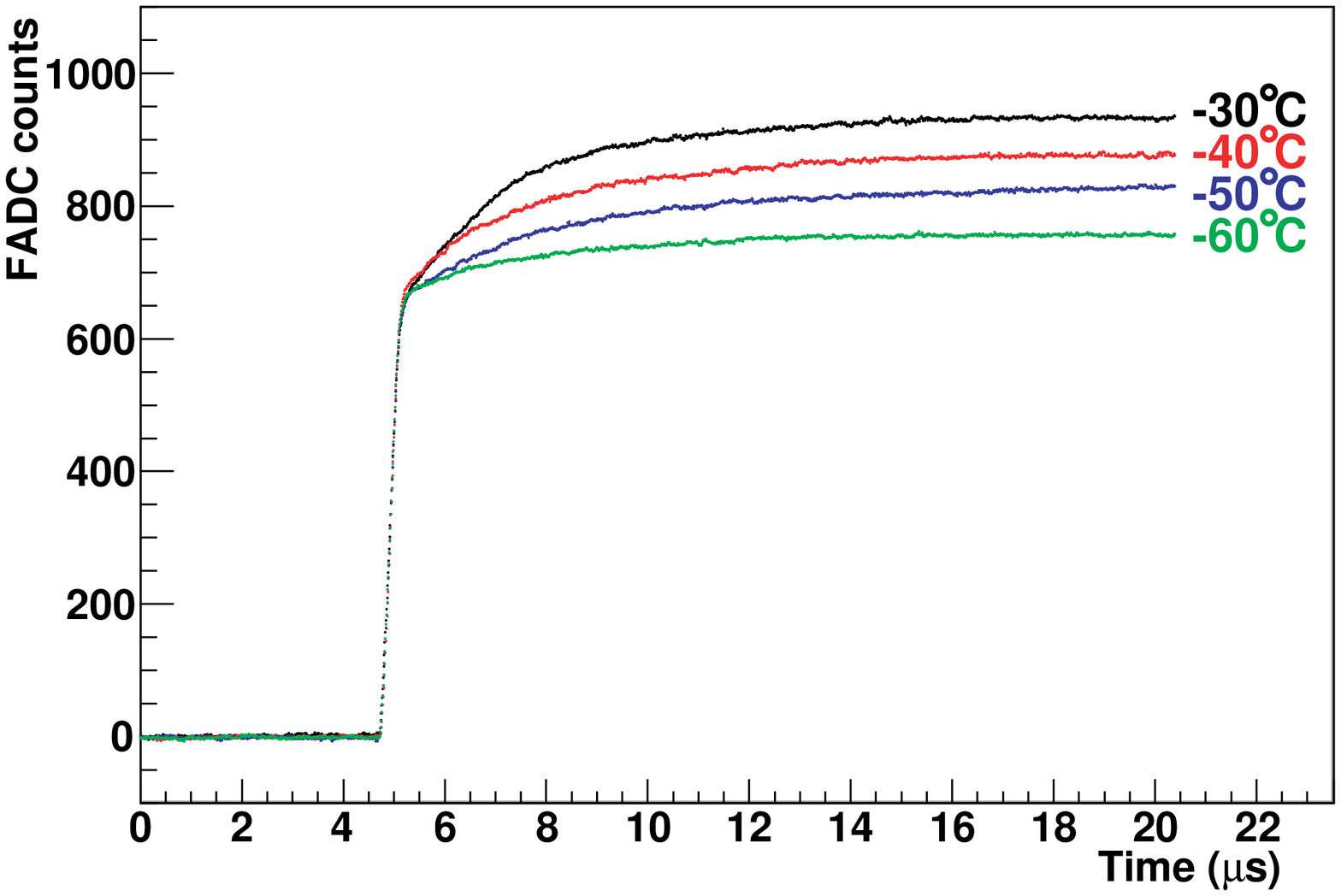}
\caption{\label{wave_temperature}Typical waveforms at various temperatures}
\end{minipage}
\end{figure}

Best energy resolution was achieved at temperatures between ${-10}^\circ\mathrm{C}$ and ${0}^\circ\mathrm{C}$.
Comparably high energy resolution was achieved
even at the room temperature (2.1\% FWHM at ${20}^\circ\mathrm{C}$)
because the temperature dependence is modest at above ${-10}^\circ\mathrm{C}$.
At high temperature, noise from leakage current and electronics becomes larger and the energy resolution becomes worse.
At low temperature, the drift time becomes longer as shown in Fig.\ref{drift_time_temperature}.
Figure \ref{wave_temperature} shows typical waveforms at various temperatures.
In this figure, events having same prompt rise height and duration were selected.
Therefore, energy deposits and positions of carrier generation are expected to be subequal.
As can be seen in this figure, the hole mobility becomes smaller
and the effect of hole trapping becomes more significant
at lower temperature.
Below ${-40}^\circ\mathrm{C}$, drift time cannot be estimated
and below ${-60}^\circ\mathrm{C}$, the effect of the hole drift disappears.
Therefore, the energy resolution becomes worse at low temperature.

\section{Conclusion}
We developed a method to achieve high energy resolution with rather thick CdTe detectors by
correcting the pulse height from the waveform of the signal
without relying on the specific models.
We demonstrated that 2.0\% (FWHM) energy resolution can be achieved
using an ohmic-contact CdTe detector with a 5.0mm $\times$ 5.0mm $\times$ 5.0mm device for 662keV $\gamma$-rays by this method.
Best energy resolution can be achieved at temperatures between ${-10}^\circ\mathrm{C}$ and ${0}^\circ\mathrm{C}$.
Comparable energy resolution was achieved even at the room temperature because the temperature dependence is modest
at above ${-10}^\circ\mathrm{C}$.

\section*{Acknowledgements}
We are grateful to Mr. K. Mori of CLEAR-PULSE Co.,Ltd for providing the CdTe detector and advice on its use.
This work was supported by Grant-in-Aid for Young Scientists B.





\bibliographystyle{model1a-num-names}

\begin{thebibliography}{00}
\bibitem{cdte_mobility}P.J.Sellin {\it et al.}, IEEE Trans. in Nucl. Sci. {\bf52}, 3074 (2005)
\bibitem{cpg}P.N.Luke, Appl. Phys. Lett. {\bf65}, 2884 (1994)
\bibitem{pixel}H.H.Barrett, J.D.Eskin, and H.B.Barber, Phys. Rev. Lett. {\bf75}, 156 (1995)
\bibitem{capture}D.S.Bale and C.Szeles, Proc. SPIE {\bf6319}, 63190B (2006)
\bibitem{system}M.Richter and P.Siffert, Nucl. Instrum. and Meth. A{\bf323}, 529 (1992)
\bibitem{bargholtz}Chr.Bargholtz, E.Fumero, L.Martensson, Nucl. Instrum. and Meth. A{\bf434}, 399 (1999)
\bibitem{takahashi}H.Takahashi {\it et al.}, Nucl. Instrum. and Meth. A{\bf458}, 375 (2002)
\bibitem{abbene}L.Abbene, G.Gerardi, Nucl. Instrum. and Meth. A{\bf654}, 340 (2011)
\bibitem{montemont}G.Montemont {\it et al.}, IEEE TRANSACTIONS ON NUCLEAR SCIENCE, {\bf52}, NO.5 (2005)
\bibitem{detrapping}M.Martini and T.A.McMath, Nucl. Instrum. and Meth. {\bf79}, 259 (1970)
\bibitem{wd}W.R.Leo, ISBN 3-540-17386-2, Springer-Verlag Heidelberg (1987)
\bibitem{fano}G.Bale, A.Holland, P.Seller and B.Lowe, Nucl. Instrum. and Meth. {\bf436}, 150 (1999)
\end{thebibliography}







\end{document}